\documentclass[review,number,sort&compress]{elsarticle} 
\usepackage[gen]{eurosym}
\usepackage{graphicx}

\usepackage{amssymb}

\journal{Nuclear Instruments and Methods A}

\begin{document}

\begin{frontmatter}



\title{The KM3NeT deep-sea neutrino telescope}

\author{Annarita Margiotta, on behalf of the KM3NeT Collaboration}
\address{INFN Sezione di Bologna and Dipartimento di Fisica e Astronomia - Universit\`a di Bologna,\\
viale C. Berti-Pichat, 6/2, 40127 - Bologna, Italy}

%
\begin{abstract}
KM3NeT is a deep-sea research infrastructure being constructed in the Mediterranean Sea. It will host the next generation Cherenkov neutrino telescope and nodes for a deep sea multidisciplinary observatory, providing oceanographers, marine biologists, and geophysicists with real time measurements. The neutrino telescope will complement IceCube in its field of view and exceed it substantially in sensitivity. Its main goal is the detection of high energy neutrinos of astrophysical origin. The detector will have a modular structure with six building blocks, each consisting of about one hundred Detection Units (DUs). Each DU will be equipped with 18 multi-PMT digital optical modules. The first phase of construction has started and shore and deep-sea infrastructures hosting the future KM3NeT detector are being prepared offshore Toulon, France and offshore Capo Passero on Sicily, Italy. The technological solutions for the neutrino detector of KM3NeT and the expected performance of the neutrino telescope are presented and discussed.
\end{abstract}

\end{frontmatter}
\section{Introduction}
\begin{figure}[htb]
\centerline{
\includegraphics[width=0.5\columnwidth]{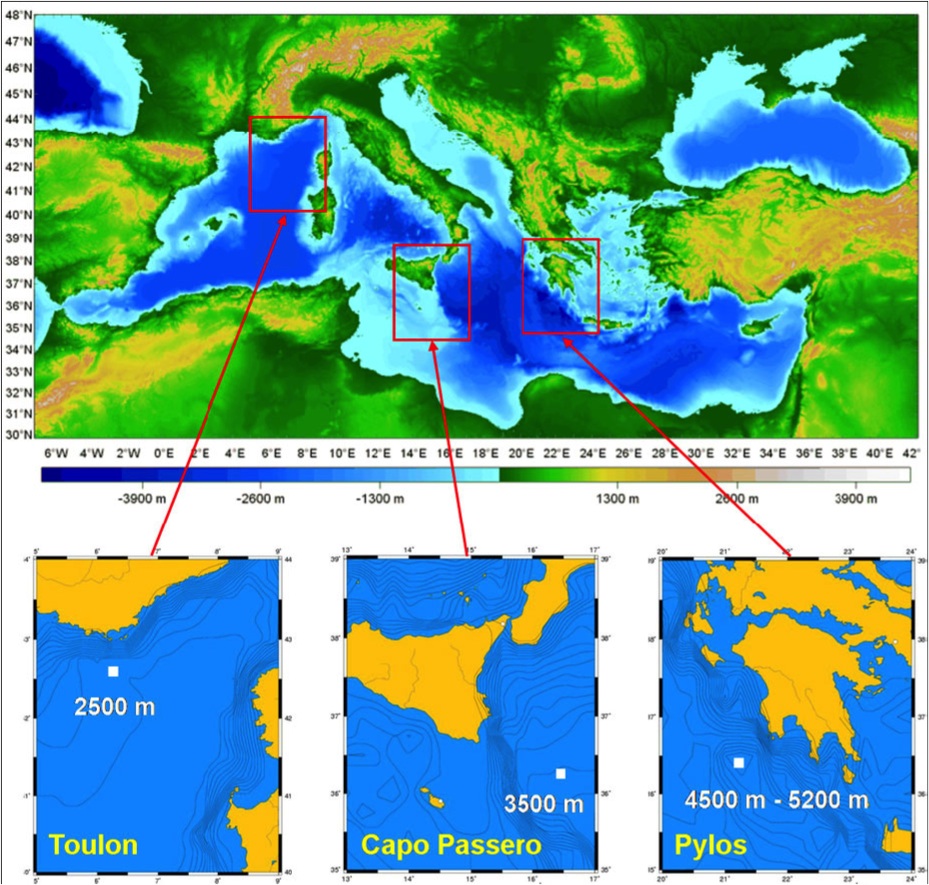}}
\caption{Location of the three proposed sites for the construction of the KM3NeT neutrino telescope in the Mediterranean Sea.}
\label{sites}
\end{figure}

The characteristics of the neutrino make it an astronomical probe that can cross extremely large distances and transport information directly from the core of its production sites.
The same features making neutrinos so useful in the exploration of the far Universe and of dense astrophysical objects put severe constraints for the construction of a neutrino detector, requiring the instrumentation of huge amount of matter.
In the 60's Markov suggested the use of ocean or lake water as target and active medium for neutrino detectors~\cite{markov}. This idea is at the basis of the neutrino telescope concept. The detection principle in a neutrino telescope relies on the measurement of the Cherenkov light emitted in a natural transparent medium, like water or ice, along the path of the particles produced in  neutrino interactions in the vicinity of a three dimensional array of photon detectors.\\ 
Due to the long muon path length, the effective size of the detector is much larger for charged current  muon neutrino interactions than for other channels. Starting from time, position and amplitude of the photon signals, dedicated algorithms can reconstruct the trajectory of the muons, inferring the neutrino direction. The neutrino and the muon directions are almost collinear at high energy, and this allows the identification of a possible source with a high resolution, better than one degree in water.
Measuring the total amount of light released in the detector, the energy of the event can, also, be evaluated with a reasonable accuracy.\\
KM3NeT is a research infrastructure housing a network of neutrino telescopes  located  in three different deep sites in the Mediterranean Sea, Fig.~\ref{sites}. 
\begin{itemize}
\item KM3NeT-Fr: this site is close to the ANTARES detector site, at a depth of about 2500 m, 20 km from the coast, south of Toulon, \cite {antares} ;
\item KM3NeT-It: this is the site chosen by the NEMO Collaboration for the deployment of a prototype tower, which has been acquiring data since March 2013, \cite {tom}. It is at 3500 m under sea level at about 100 km from Capo Passero, in Sicily;
\item KM3NeT-Gr: this is the deepest site of the three proposed locations, at 4500/5000 m under sea level, at about 30 km southwest of Pylos, in Greece.  
\end{itemize}
The design of the KM3NeT telescope is optimised for the search for Galactic neutrino sources. With its location in the Mediterranean Sea, the Galactic centre is visible for about 75\% of the time. The excellent optical properties of water allow for a good pointing accuracy.
The building concept of the telescope foresees the construction of several large detector blocks to be installed in  different sites. The cost of the complete research infrastructure, which will  also host   a multidisciplinary observatory for Earth and deep sea science, is estimated around 220-250 MEuros.\\
The first phase of the KM3NeT infrastructure is being installed at the KM3NeT-Fr  and KM3NeT-It  sites, using the present total available funding (31 M\euro{}). The full detector will also include  the  Km3NeT-Gr site.
\begin{figure}[htb]
\centerline{
\includegraphics[width=0.5\columnwidth]{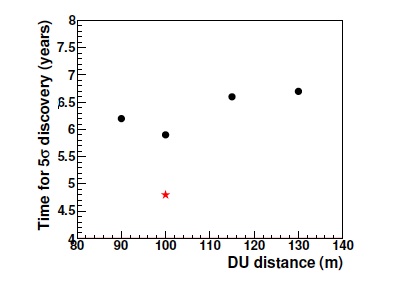}}
\caption{Years for discovery at 5$\sigma$s, 50\% probability as a function of the DU distance for the RXJ1713.
The black dots are the values calculated with a binned method and the red star is the unbinned value
extrapolated from one year flux.}
\label{rxj}
\end{figure}
\section{The physics case}
The main goal of the KM3NeT  telescope is the discovery of high-energy neutrino sources. Its location is optimal to look directly at the Galactic centre, which  hosts several TeV gamma-ray sources that, in a hadronic  scenario, could emit measurable neutrino fluxes. The excellent optical properties of water (scattering and absorption lengths) allow the reconstruction of muon tracks with an angular resolution better than one degree in the energy range where a signal is expected (a few TeV to a few 10 TeV). This makes the charged current interactions of muon neutrinos in the vicinity of the detector the golden channel for source identification.
The most effective strategy to identify neutrino induced muons is the simple geometrical selection of upward going tracks. Nevertheless, other channels, like neutral current interactions, will be explored as well.

An intense work of simulation has been carried out in order to optimise the detector design for the discovery of Galactic sources.
In particular, the performance of KM3NeT for detection of neutrinos from the supernova remnant RXJ1713.7-3946 and the pulsar wind nebula Vela X have been studied. Gamma rays in the TeV region, which can be explained with the hypothesis of hadronic mechanisms acting inside these objects, have been measured.   Assuming an exponential cut-off power law for the neutrino energy spectrum and simulating different solutions for the detector design, the conclusion is that with a detector arrangement as described in section 2, KM3NeT can claim a discovery  after about 5 years of observation, Fig.~\ref{rxj}, and the observation at a significance level of 3$\sigma$  with 50\%CL  after 2 years, for  RXJ1713.7-3946. A shorter observation time (about 3 years for the discovery and slightly more than 1 year for the observation) would be necessary in the case of Vela X, \cite{agata}.\\
Data from the  Fermi Large Area Telescope  showed evidence for the emission of high energy gamma rays from two large areas, in the region above and below the Galactic Centre (the so-called "Fermi bubbles" \cite{fb-antares}. Assuming  a hadronic mechanism for the gamma ray production from the Fermi bubbles, a high energy neutrino flux is also expected. If  the gamma production is completely due to  hadronic processes, the results of Monte Carlo simulations \cite{fb-km3net} indicate that a discovery is possible in about one year of KM3NeT operation.\\
Accurate simulations have shown that partitioning the detector in blocks of about half km$^3$ provides the optimal sensitivity for detection of neutrino sources \cite {multisite}. This concept allows for phased construction and a larger technical reliability of the detector. In addition, the distribution of the blocks over different locations in the Mediterranean maximises the possibility of regional funding. \\
Although KM3NeT Phase-1 and Phase-2 detectors have been optimised to the search for neutrinos sources, the recent results of the IceCube experiment~\cite{icecube} have prompted the KM3NeT Collaboration to consider also an intermediate KM3NeT Phase 1.5, whose main goal is to measure a diffuse flux neutrino signal, using different methodology, with complementary field of view and with a better angular resolution. The Phase1.5 infrastructure will comprise 2 building blocks to be realized at the KM3NeT-Fr and KM3NeT-It sites through an update of the existing installation. The additional estimated cost is about 50-60 M\euro{}. Sensitivity studies are in progress.
\begin{figure}[htb]
\centerline{
\includegraphics[width=0.5\columnwidth]{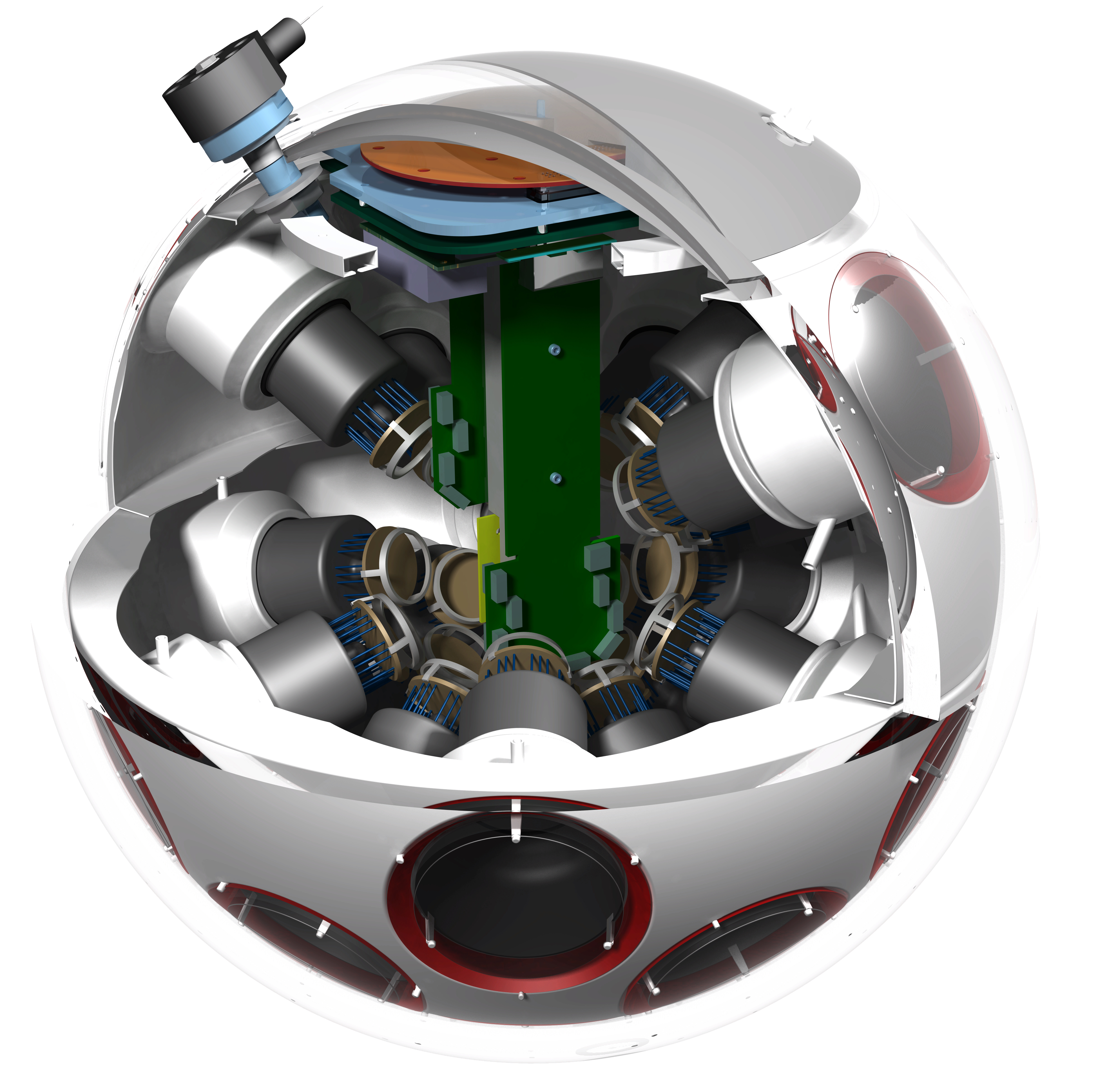}}
\caption{Internal structure of a  KM3NeT DOM.}
\label{internal}
\end{figure}
\begin{figure}[htb]
\centerline{
\includegraphics[width=0.5\columnwidth]{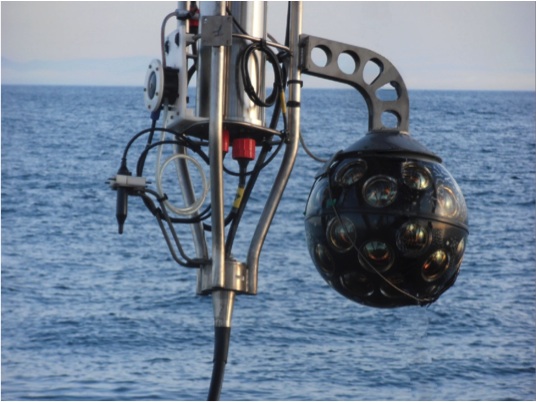}}
\caption{Photo of the prototype multiPMT DOM, presently in data taking on the ANTARES detector.}
\label{ppmdom}
\end{figure}
\section{The detector.}
\begin{figure}[htb]
\centerline{
\includegraphics[width=0.5\columnwidth]{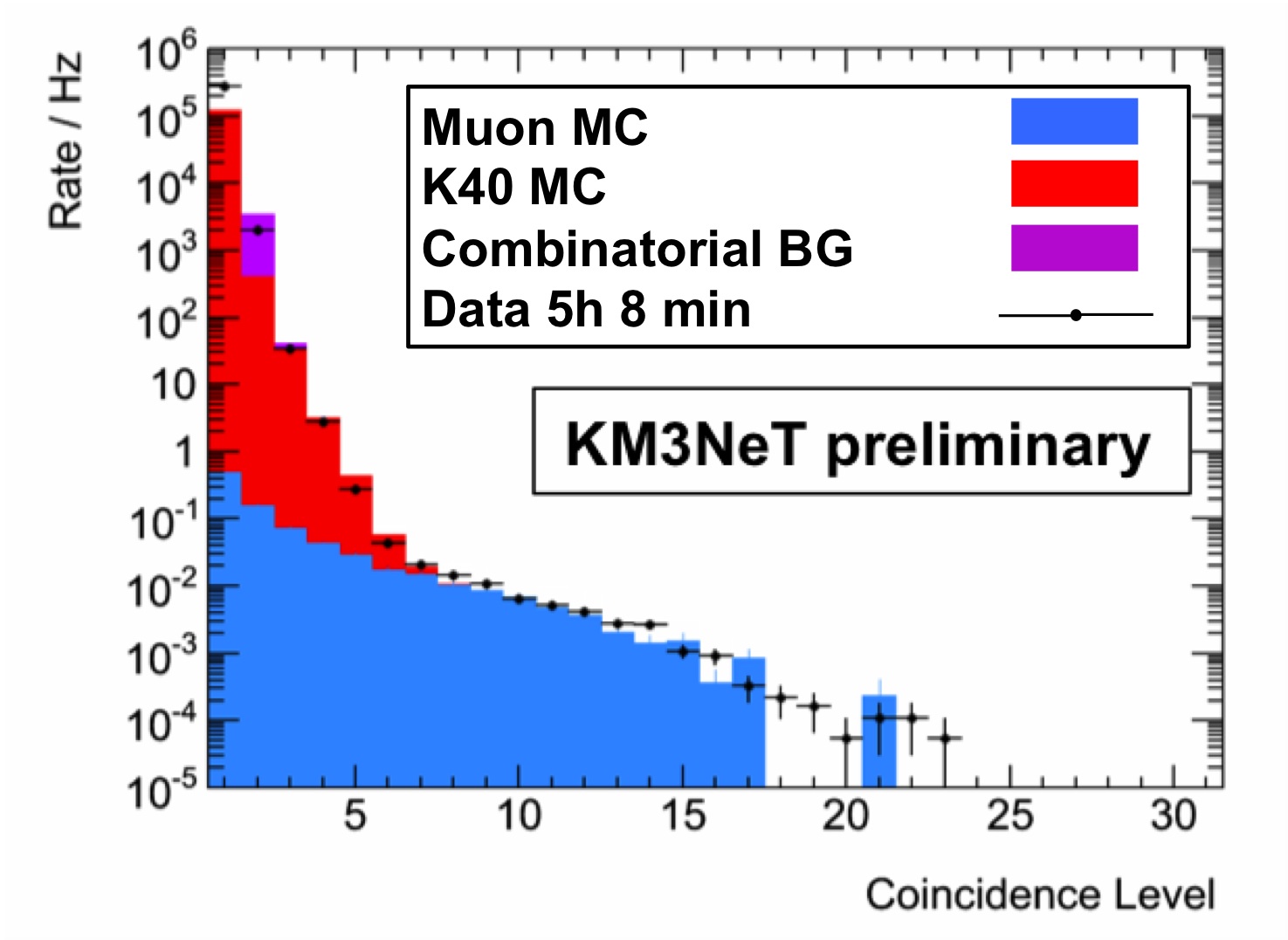}}
\caption{Comparison between the measured and expected rate on the PPM-DOM hosted by the ANTARES detector as a function of the number of coincident hits. The red histogram represents the contribution due to the decay of $^{40}$K dissolved in the sea water. Measurements are still ongoing and the plot must be considered as preliminary.}
\label{rate}
\end{figure}

Although the ANTARES technological choices have been the starting point, several years of R\&D efforts have produced new technological solutions for the KM3NeT detector.\\
A major innovation concerns the design of the active part of a neutrino telescope, the optical module. The digital optical module (DOM) used in KM3NeT is a 17-inch glass sphere, resistant to the high pressure present at the sea bottom,  housing 31 3-inch photomultiplier tubes (PMT),  the active bases for power and the readout electronics. The bases  can be controlled individually from the shore in order to set the correct values of HV and threshold for each tube. The support structure for the PMTs is constructed using 3D-printing technology. Optical gel fills the cavities to ensure optical contact.
A metallic structure glued on the sphere supports the electronic boards for their cooling, Fig.~\ref{internal}. The maximum expected power consumption of a DOM is 10 W.\\
A multiPMT DOM has several advantages if compared to the traditional, large cathode single-PMT optical module used in ANTARES, Baikal and IceCube detectors. In particular, it has a larger (three to four times) total photocathode area and a better discrimination of single vs multi photoelectron. A light collection ring around the front of each PMT further increases the effective photocathode area by 20-40\%. \\
\begin{table}[htdp]
\caption{Main characteristics required for the 3-inch PMT}
\begin{center}
\begin{tabular}[width=0.5\columnwidth]{lc}
\hline
Photocathode's diameter & $>$72 mm\\
Dynodes & 10 \\
Nominal Voltage for Gain $3\cdot 10^6$ & 900 - 1300 V \\
Gain slope & 6.5 min -8.0 max\\
QE at 404 nm &  $>$ 23 \%\\
QE at 470 nm & $>$ 18 \%\\
Collection efficiency & $>$ 87\% \\
Uniformity of QE and Coll. eff. & within 20\%\\
TTS (FWHM) & $<$ 5 ns \\
Dark count rate (0.3 s.p.e. threshold) &   $< $ 2ÊkHz\\
Pre-pulses &   $< $ 1 \% \\
Delayed pulses & $< $ 3.5 \% \\
Early afterpulses & $< $ 2 \% \\
Late afterpulses &  $< $ 10 \% \\
\hline
\end{tabular}
\end{center}
\label{default}
\end{table}
Three companies - ETEL, Hamamatsu and HZC - have developed new PMT types, according to the main specifications required for the KM3NeT detector, see Table 1.
For the first phase of KM3NeT, Hamamatsu has already started to deliver PMTs. Negotiations with ETEL and HZC are not yet concluded.\\
Each DOM acts independently as an IP node. All DOMs are synchronised to the subnanosecond level using a clock signal broadcast from shore. Monitoring and real-time correction for the propagation delays between the shore station and each single DOM will be performed using a White Rabbit application,~\cite{wr}.
Calibration sensors are also contained in the optical modules: light beacons to illuminate groups of DOMs at known times in order to monitor individual time offsets, piezo sensors for acoustic positioning, a tiltmeter, a compass and sensors for temperature and humidity measurements inside the sphere.
The readout electronic boards (Central Logic Board, CLB), controlling data acquisition and communications with the shore station, are also hosted in the DOM.
The signal from a PMT  consists of the arrival time and the width of the pulse, measured as the time-over-threshold, typically set at 0.3 p.e., digitised and sent via a network of optical fibers to shore. Long range transmissions exploit DWDM techniques at 50 GHz spacing.\\  The "all-data-to-shore" concept is applied to the readout of the detector, following the experience of the ANTARES detector. On shore, the physics events are filtered from the background using a dedicated software and are stored on disk.\\
A prototype DOM has been installed on the ANTARES detector (Fig.~\ref{ppmdom})  in the spring of 2013 and has been taking data since then, according to expectations.  Fig.~\ref{rate} shows  a preliminary comparison between collected data and MC expectations. Simply requiring a coincidence among 6-7 PMTs   the bulk of hits due to the decay of $^{40}$K dissolved in the sea water can  be rejected. Measurements are still on going.\\
\begin{figure}[htb]
\centerline{
\includegraphics[width=0.5\columnwidth]{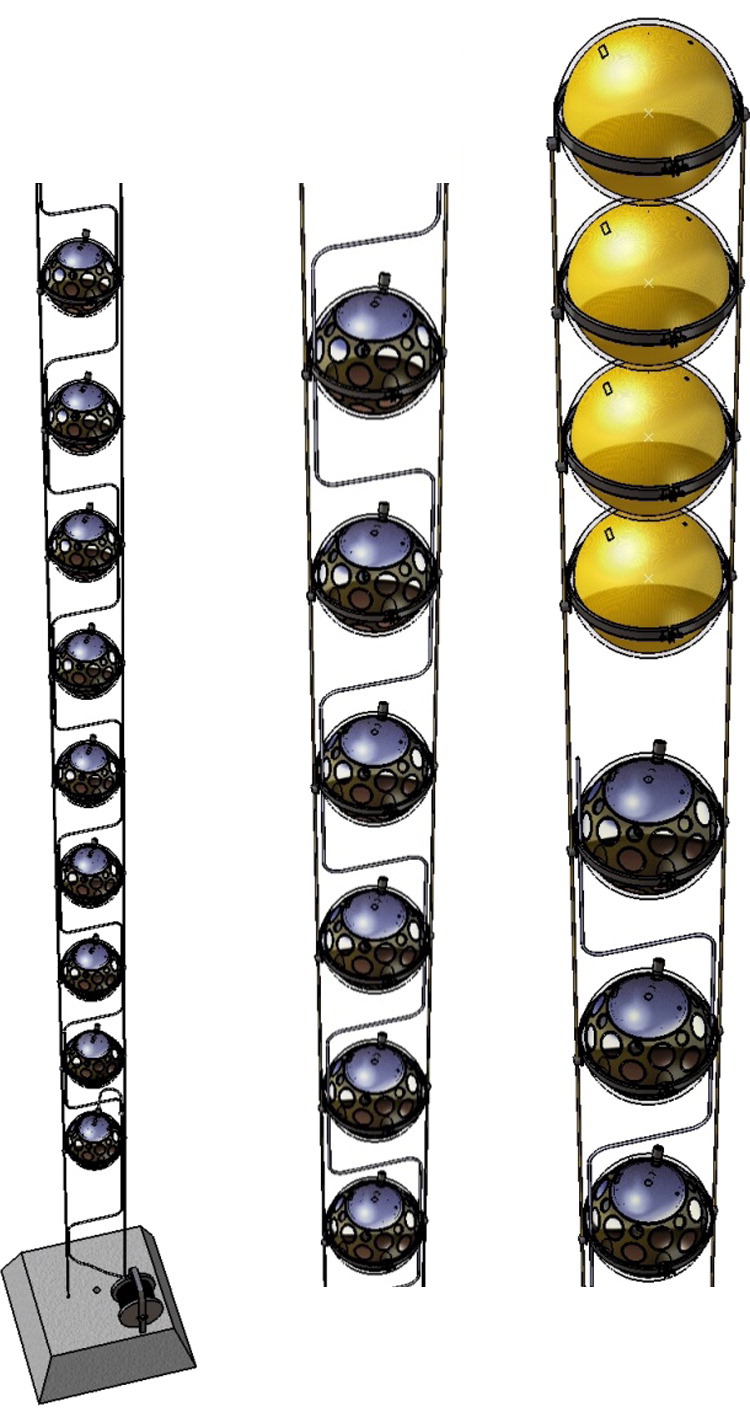}}
\caption{Schematic view of a detection unit of KM3NeT. The yellow spheres represent buoys, the dark ones are the DOMs.}
\label{string}
\end{figure}
18 DOMs are attached along two thin parallel Dyneema ropes with a vertical separation of 36m between DOMs.  
A vertical electro-optical cable, an oil filled plastic tube containing  copper wires and  optical fibres for power and data transmission, is also attached to the ropes. This flexible string, about 700 m long, is the main mechanical structure of a detection unit (DU), Fig.~\ref{string}. A system of buoys keeps the string taut. The base anchors the string on the sea bottom and houses a power converter and dedicated optical components. It also  contains the interlink cable connecting each string to the rest of the building block, made of 115 strings at a distance of 90-100 m from each other, for a total instrumented volume of about half km$^3$/building block. A main electro-optical cable connects the deep sea infrastructure to the shore station for power distribution and data transmission. 
\begin{figure}[htb]
\centerline{
\includegraphics[width=0.5\columnwidth]{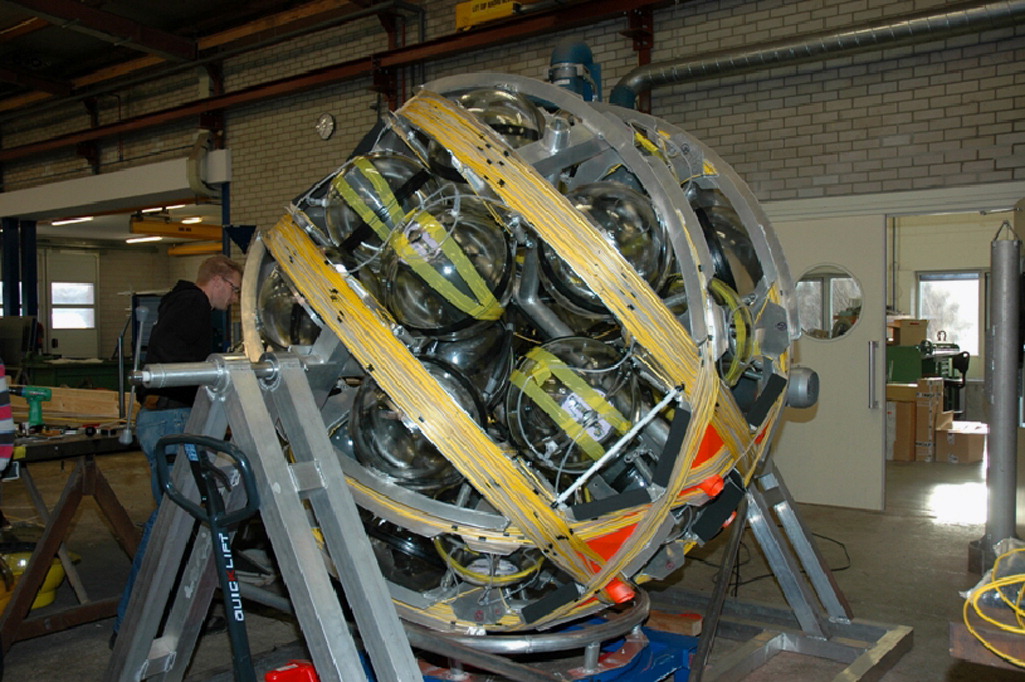}}
\caption{Photo of the launcher vehicle for optical modules (LOM) .}
\label{lom}
\end{figure}
\begin{figure}[htb]
\centerline{
\includegraphics[width=0.5\columnwidth]{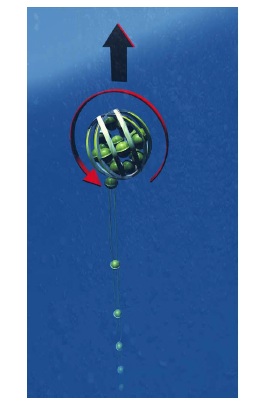}}
\caption{Unfurling a DU of KM3NeT. Courtesy Marijn van de Meer/Quest.}
\label{unroll}
\end{figure}
The final complete configuration (KM3NeT Phase 2) foresees six building blocks with each site hosting at least one building block.\\
The deployment of such a large number of DU requires a special technique, ~\cite{lom-pap}. A recoverable launching vehicle has been designed and successfully tested during several cruises. Before deployment, each string is wound on a spherical aluminum launching vehicle, the LOM,  Fig.~\ref{lom}. From the vessel, the LOM is lowered to the seabed and an acoustic release allows the launcher to freely float to the surface, where it can be easily recovered, while unrolling the string, Fig.~\ref{unroll}.
The KM3NeT Phase 1, which is completely funded, will see the construction and deployment of 31 DUs, 7 at the Toulon site and 24 at Capo Passero. Here a set of 8 "towers" built according to the old NEMO Collaboration design will also be deployed, starting at the end of 2014, and independently operated. \\
\section{Conclusions}
The first phase of the KM3NeT neutrino telescope has started. A prototype of the digital optical module specifically designed for this new apparatus   was installed on the ANTARES detector and has been taking data according to the expectations.
The project of the telescope has been optimised to search for Galactic neutrino sources. Simulations show that in a few years of observation evidence of a neutrino flux from the Fermi bubbles or from some astrophysical objects like supernova remnants or pulsar wind nebulae can be reached even with a partially complete detector.
Studies are in progress to estimate the sensitivity of KM3NeT to a neutrino diffuse flux (IceCube-like events) and to consider a possible re-optimization of the detector to this signal.

\end{document}